\documentclass{iopconfser}
\usepackage{orcidlink}
\usepackage{multirow,ragged2e}
\usepackage{xcolor,colortbl}

\usepackage[numbers,sort&compress]{natbib}

\usepackage{acronym}
\acrodef{GW}[GW]{gravitational wave}
\acrodef{GWECS}[GWECS]{Gravitational-Wave Early Career Scientists}
\acrodef{ECR}[ECR]{early-career researcher}
\acrodef{Amaldi}[Amaldi]{Edoardo Amaldi Conference on Gravitational Waves}
\acrodef{GR}[GR]{International Conference on General Relativity and Gravitation}
\acrodef{LISA}[\textit{LISA}]{\textit{Laser Interferometer Space Antenna}}
\acrodef{PTA}[PTA]{pulsar timing array}
\acrodef{ESA}[ESA]{European Space Agency}

\usepackage{etoolbox}
\newtoggle{checklength}
\toggletrue{checklength} 

\begin{document}

\title{The Early Career Workshop of GR--Amaldi 2025}

\author{S~Al-Shammari$^{1}$\orcidlink{0009-0006-6367-2332},
C~P~L~Berry$^{2}$\orcidlink{0000-0003-3870-7215}, 
C~E~A~Chapman-Bird$^{3}$\orcidlink{0000-0002-2728-9612}, 
F~T~Chowdhury$^{4,5}$\orcidlink{0000-0001-8229-2374}, 
K~Cunningham$^{6}$\orcidlink{0000-0002-1094-2267},
M~Emma$^{7}$\orcidlink{0000-0001-7943-0262},
R~Gray$^{2}$\orcidlink{0000-0002-5556-9873}, 
C~Hoy$^{8}$\orcidlink{0000-0002-8843-6719}, 
I~S~Heng$^{2}$\orcidlink{0000-0002-1977-0019},
M~Korobko$^{9}$\orcidlink{0000-0002-3839-3909},
E~Maggio$^{10}$\orcidlink{0000-0002-1960-8185}, 
A-K~Malz$^{7}$\orcidlink{0009-0004-7196-4170},
H~Middleton$^{3}$\orcidlink{0000-0001-5532-3622},
M~Prathaban$^{11}$\orcidlink{0009-0004-1672-4086}, 
I~M~Romero-Shaw$^{1,11,12}$\orcidlink{0000-0002-4181-8090},
G~Shaifullah$^{13,14,15}$\orcidlink{0000-0002-8452-4834}, 
S~Singh$^{2}$\orcidlink{0000-0003-4881-1067},
J~Steinlechner$^{16,17}$\orcidlink{0000-0002-6697-9026}, 
K~Toland$^{2}$\orcidlink{0000-0001-9537-9698},
D~Williams$^{2}$\orcidlink{0000-0003-3772-198X},
and
M~J~Williams$^{8}$\orcidlink{0000-0003-2198-2974}
}

\iftoggle{checklength}{
\affil{$^{1}$Gravity Exploration Institute, Cardiff University, Cardiff, United Kingdom}

\affil{$^{2}$Institute for Gravitational Research, University of Glasgow, Kelvin Building, University Ave., Glasgow, G12 8QQ, United Kingdom} 

\affil{$^{3}$Institute for Gravitational Wave Astronomy, University of Birmingham, Birmingham, B15 2TT, United Kingdom}

\affil{$^{4}$Department of Physics and Astronomy, University of Exeter, Exeter EX4 4QL, United Kingdom}

\affil{$^{5}$Department of Chemistry, The Ohio State University, Columbus, Ohio 43210, United States}

\affil{$^{6}$School of Mathematics \& Statistics, University College Dublin, Belfield, Dublin 4, Ireland}

\affil{$^{7}$Department of Physics, Royal Holloway, Egham, TW20 0EX, United Kingdom}

\affil{$^{8}$Institute of Cosmology and Gravitation, University of Portsmouth, Portsmouth, United Kingdom}

\affil{$^{9}$Institut für Quantenphysik and Zentrum für Optische Quantentechnologien, Universität Hamburg, Luruper Chaussee 149, 22761 Hamburg, Germany}

\affil{$^{10}$Max Planck Institute for Gravitational Physics (Albert Einstein Institute), Am Mühlenberg 1, 14476 Potsdam, Germany}

\affil{$^{11}$Kavli Institute for Cosmology, University of Cambridge, Cambridge, CB3 0HA, United Kingdom}

\affil{$^{12}$H.H. Wills Physics Laboratory, University of Bristol, Tyndall Avenue, Bristol BS8 1TL, United Kingdom}

\affil{$^{13}$Dipartimento di Fisica ``G. Occhialini'', Università degli Studi di Milano-Bicocca, Piazza della Scienza 3, I-20126 Milano, Italy}

\affil{$^{14}$INFN, Sezione di Milano-Bicocca, Piazza della Scienza 3, 20126 Milano, Italy.}

\affil{$^{15}$INAF - Osservatorio Astronomico di Cagliari, via della Scienza 5, 09047 Selargius (CA), Italy.}

\affil{$^{16}$Maastricht University, Minderbroedersberg 4-6, 6211 LK Maastricht, The Netherlands}

\affil{$^{17}$Nikhef, Science Park 105, 1098 XG Amsterdam, The Netherlands}
}

\email{christopher.berry.2@glasgow.ac.uk}

\newpage
\begin{abstract}
Gravitational physics and astronomy have developed rapidly over the last decade, driven by new observations and theoretical breakthroughs. 
As new as the science and technology of this field are, its greatest asset may be the body of \aclp{ECR} actively engaged in driving it forward. 
With the aim of bringing together this community of enthusiastic scientists from a broad array of disciplines, the organisers of the \acs{GR}--\acs{Amaldi} meeting joined with the \acl{GWECS} organisation to create a three-day event---the Early Career Workshop. 
The Workshop aimed to provide a broad overview of the field’s diverse scientific possibilities and introduce key theoretical foundations underpinning its science. 
To complement developing technical skills, the Workshop also sought to provide participants with transferable skills to aid them in their future careers. 
The Workshop emphasized networking and community building, offering participants opportunities to engage with peers and mentors.
It encouraged interdisciplinary exchanges and cross-institutional collaboration, fostering connections across different research efforts. 
Collectively, these initiatives aimed to equip participants with a comprehensive understanding of the field's research and to build a more cohesive, collaborative community of \aclp{ECR}. 
We summarise key points and conclusion from the various activities carried out as part of the Workshop.
\end{abstract}

\section{Youthful science: The need for an Early Career Workshop}

Although modern gravitational science, incorporating general relativity and \acp{GW}, has been established for over 100 years, recent observations have sparked a rapid growth in its community.
The first direct observation of \acp{GW}~\cite{LIGOScientific:2016aoc}
not only verified directly Einstein's predictions~\cite{LIGOScientific:2016lio,LIGOScientific:2021sio}, but also highlighted the promise of many exciting discoveries in the years to come~\cite{LIGOScientific:2025hdt,KAGRA:2021vkt,LIGOScientific:2025slb}. 
At the same time, advances in computing and the rise of big data allowed for conceptually new approaches in analysing the data and modelling complicated cosmic systems~\cite{LSST:2008ijt,Zhang:2015,Allen:2019dkq,EventHorizonTelescope:2019uob,Dvorkin:2022pwo,Cuoco:2024cdk}. 
This has brought many young scientists to the field of gravity, who share not only the passion for science, but also the desire to develop skills for their future careers. 
This field has become a unique ground for uniting scientists across a broad range of topics and building a strong community for the upcoming decades of scientific breakthroughs.

Beyond the ground-based \ac{GW} detectors that made the first observations~\cite{LIGOScientific:2014pky,VIRGO:2014yos,KAGRA:2018plz}, the field has expanded rapidly through complementary probes. 
The planned \ac{LISA} mission~\cite{LISA:2024hlh} has now been adopted by the \acl{ESA}, marking a decisive step towards opening the low-frequency \ac{GW} window from space~\cite{LISA:2022yao,LISACosmologyWorkingGroup:2022jok,LISA:2022kgy}. 
\Acl{PTA} collaborations have recently announced evidence for a stochastic \ac{GW} background~\cite{NANOGrav:2023gor,EPTA:2023fyk,Reardon:2023gzh,Xu:2023wog}, likely generated by a cosmic population of supermassive black hole binaries~\cite{InternationalPulsarTimingArray:2023mzf}, offering another independent view of the \ac{GW} sky. 
Parallel developments in other areas of gravitational physics further illustrate the breadth of the field: the Event Horizon Telescope has resolved black hole shadows \cite{EventHorizonTelescope:2019dse,EventHorizonTelescope:2022wkp}, surveys like \textit{Euclid} are beginning to map large-scale cosmic structure with unprecedented precision \cite{Euclid:2025rvk}, and multimessenger campaigns continue to tie gravitational physics to astrophysics and cosmology in ever more direct ways~\cite{LIGOScientific:2017adf,Margutti:2020xbo,Engel:2022yig,Nicholl:2024ttg}. 
Together, these advances highlight the interdisciplinary character of the field and the need to prepare \acp{ECR} for careers that span across traditional disciplinary boundaries.

The \ac{GWECS} initiative was established to address exactly these needs: to provide \acp{ECR} with training, community, and visibility within large international collaborations.%
\footnote{\ac{GWECS} webpage \href{http://gwecs.org}{gwecs.org}.}
Building a career in gravitational physics today almost inevitably involves working in large, distributed teams, navigating authorship and credit in publications with thousands of co-authors, and developing transferable skills in project management, communication and collaboration. 
These are challenges not unique to gravitational physics~\cite{Smith:2016,Battiston:2019}, but particularly acute in this field, given the scale of current and future projects. 
In the context of the increasingly competitive academic environment, with limited permanent positions and growing reliance on international mobility, the demands placed upon \acp{ECR}~\cite{Janssens:2023ufo,Allen:2024lyp} makes it critical to support them through structured opportunities such as workshops. 
Such initiatives not only strengthen the individual careers of young scientists but also reinforce the long-term health and inclusivity of the research community as a whole.

\subsection{Motivating opportunity}

The \ac{GR} is held every three years and is the principal international meeting for scientists working in gravity and relativity. 
The \ac{Amaldi} is held every two years and is the principal international meeting for scientists working in \ac{GW} science. 
Every six years, the \ac{GR} and \ac{Amaldi} meetings are held together; this joint meeting brings together experts from across classical and quantum gravity, mathematical and applied relativity, \ac{GW} instrumentation and data analysis, multi-messenger astronomy, relativistic astrophysics and cosmology, as well as education, outreach and community building in support of these areas. 
Conference attendance is an important part of academic career development~\cite{Sanders:2022,Agarwal:2022gno}. 
The gap between joint \ac{GR}--\ac{Amaldi} meetings makes them a rare opportunity for \acp{ECR}: most PhD students may only attend one during the course of their doctoral studies.
Given the significance of the joint meeting, and the lack of in-person meetings over the last few years, the \ac{GR}--\ac{Amaldi} meeting in Glasgow was identified as an important opportunity to bring together \acp{ECR} in the field.

The size of the \ac{GR}--\ac{Amaldi} meeting means that a large number of \acp{ECR} were expected to attend. 
At the meeting, many \acp{ECR} would present their work, attend presentations on topics outside of their main area, and interact with senior figures in the field. 
This makes the meeting an ideal setting for an Early Career Workshop as well as highlighting the benefits of a Workshop:
Technical sessions would help orient attendees ahead of the diverse program of the main meeting, transferable-skills sessions would prepare attendees to make the most out of the main meeting, and networking opportunities would help them build connections to avoid feeling lost in the crowd of hundreds of attendees of the main meeting.

To ensure the Workshop was as accessible as possible, it was important to keep it economical~\cite{Agarwal:2022gno}. 
Being attached to the \ac{GR}--\ac{Amaldi} meeting meant that there would be no additional travel needed for attendees~\cite{Gokus:2024}. 
However, there would still be subsistence and accommodation costs, as well as registration for the \ac{GR}--\ac{Amaldi} meeting. 
It was decided that the Workshop would have no additional registration fee. 
Meeting costs were covered through sponsorship. 
This was sufficient to cover catering during the breaks, plus lunch on Thursday and Friday. 
Low-cost accommodation at the University of Glasgow was provided for the duration of the Workshop and \ac{GR}--\ac{Amaldi} meeting, and Workshop attendees were given priority in booking. 
Unfortunately, it was not possible to cover travel expenses for presenters. 
However, since many were attending the \ac{GR}--\ac{Amaldi} meeting anyway, this did not pose a major difficulty, and some presenters even agreed to travel to Glasgow despite not attending the main meeting. 

\subsection{Organising committee}

The Workshop's organising committee consisted of members of the \ac{GR}--\ac{Amaldi} local organising committee and the \ac{GWECS} committee, as well as an early-career representative of the sponsoring Institute of Physics Mathematical \& Theoretical Physics Group. 
The \ac{GR}--\ac{Amaldi} local organising committee was formed by members of academic staff from the hosting University of Glasgow, who then invited \acp{ECR} from institutions across the United Kingdom and Ireland to join. 
Subcommittees were formed covering Outreach \& Media (which primarily organised public-engagement activities), Glasgow \& Family (which primarily organised resources about facilities in the city and arrangements for conference delegates bringing children), and \acp{ECR} \& Diversity (which primarily oversaw plans to make the meeting welcoming, inclusive and accessible, including organising the Community Best Practices Lunch). 
Members of the \acp{ECR} \& Diversity subcommittee joined the organising committee for the Workshop. 
Having a high proportion of \acp{ECR} on the organising committee helped to centre the meeting around the needs and priorities of current \acp{ECR} in the field.

\section{Workshop programme}

The Workshop spanned three days.
The first two days featured the core programme: a mix of talks and panel discussions.
As with the previous \ac{GWECS} workshop~\cite{Bayle:2021ylp}, the schedule was split roughly evenly between technical topics and transferable skills. These sessions were deliberately interleaved, so participants could still engage with a balanced programme even if they only attended part of it.
The third day was dedicated to networking and social activities, aimed at consolidating the connections built during the earlier sessions. The complete timetable is given in Table~\ref{tab:timetable}.%
\footnote{On Sunday 13 July (between the Workshop and the main meeting) there were several satellite events.
Attendees were invited to a Wikipedia Edit-a-thon, which provided training on editing Wikipedia and included collaborative work updating pages relevant to the meeting. 
In the evening, Erin Macdonald (a Workshop panellist) delivered a public lecture, \emph{Spacetime in Star Trek}, combining reflections on her career with discussion of how scientific concepts are represented in popular media.}
\begin{table}
    \centering
    \begin{tabular}{c p{4.8cm} p{4.8cm} p{2.9cm} }\hline
        {  } & \multicolumn{3}{c}{Day} \\\cline{2-4} 
        Time & \multicolumn{1}{c}{Thursday 10 July} & \multicolumn{1}{c}{Friday 11 July} & \multicolumn{1}{c}{Saturday 12 July} \\
        \hline\hline
        09:00--09:30 & Welcome and advertising of opportunities & Introduction to electromagnetic observations of relativistic systems (Antonio Carrillo) & \multirow[t]{15}{=}{Social networking (Pollock Country Park \& Burrell Collection; Glasgow West End highlights;	Ben Bowie \& Loch Lomond)} \\
        09:30--10:00 & \multirow[t]{2}{=}{Ice-breaking and networking} & Introduction to cosmological surveys (Andy Taylor) & \\
        10:00--10:30 &  & Q\&A & \\
        \cline{1-3}
        10:30--11:00 & \multicolumn{2}{c}{\textit{Morning coffee break}} & \\
        \cline{1-3}
        11:00--11:30 & Introduction to black hole spacetimes (Scott Hughes) & \multirow[t]{3}{=}{Careers in academia panel (Anna Green, Simone Mastrogiovanni, Christiana Pantelidou, Adam Pound)} & \\
        11:30--12:00 & Introduction to solving Einstein’s equations (Katy Clough) & & \\
        12:00--12:30 & Q\&A & & \\
        \cline{1-3}
        12:30--13:30 & \multicolumn{2}{c}{\textit{Lunch break}} & \\
        \cline{1-3}
        13:30--14:00 & Introduction to public engagement (Jen Gupta) & Introduction to alternative theories of gravity (Daniela Doneva) & \\
        14:00--14:30 & Introduction to attending conferences (Mikhail \mbox{Korobko}) & Introduction to observational tests of general relativity (Elisa Maggio) & \\
        14:30--15:00 & Introduction to academic writing (Christopher Berry) & Introduction to \ac{GW} experiments (Jonathan Gair) \\
        15:00-15:30 & Q\&A & Q\&A & \\
        \cline{1-3}
        15:30--16:00 & \multicolumn{2}{c}{\textit{Afternoon tea break}} & \\
        \cline{1-3}
        16:00--16:30 & Introduction to quantum theories of gravity (Bianca Dittrich) & \multirow[t]{3}{=}{Careers outside academia panel (Rebecca Douglas, Sebastian Khan, Erin Macdonald, Jen Toher, Peter Wakeford)} & \\
        16:30--17:00 & Introduction to data analysis (John Veitch) & & \\
        17:00--17:30 & Q\&A &  & \\
        \cline{1-4}
        Evening & \textit{Free} & Pub quiz & \textit{Free} \\
        \hline
    \end{tabular}
    \caption{
    Workshop schedule and invited presenters. 
    }
    \label{tab:timetable}
\end{table}

\subsection{Networking}

Day 1 opened with a networking and ice-breaking session designed to spark conversations and build connections.  
The first activity, \emph{People Bingo}, invited participants to circulate and match 25 listed traits (e.g., ``first in-person conference attendee'' or ``primarily codes in a language other than Python'') to others in the room. 
This was followed by group \emph{Two Truths and a Lie} exercises, submitted via the online Mentimeter interactive-presentation platform~\cite{Mohin:2022,Khan:2025} for the room to vote on. 
Finally, groups suggested entries for a \emph{\ac{GR}--\ac{Amaldi} Bingo Card}, with the best to be compiled and shared at the main conference.  

The exercises kept the atmosphere lively and collaborative. Several participants completed their Bingo cards within 15 minutes, and groups enthusiastically submitted ideas. 
Mentimeter encouraged close group discussion while seamlessly feeding results back, which helped establish a sense of cohesion early on. 
Overall, the session succeeded in setting an engaging and positive tone for the Workshop~\cite{Chlup:2010,Ravn:2011}.

\subsection{Lectures}

Given the multidisciplinary scope of the \ac{GR}--\ac{Amaldi} meetings, the workshop included lectures introducing the main topics and providing entry points for discussion. 
Experienced researchers presented broad overviews of their fields, beginning with accessible introductions and leading into current frontiers. 
In total, nine lectures covered scientific and technical topics, and three focused on transferable skills.  

The opening lecture, by Scott Hughes (MIT), introduced black hole spacetimes, covering their unique theoretical properties~\cite{Israel:1967wq,Carter:1971zc,Robinson:1975bv}, observables~\cite{Ghez:2008ms,GRAVITY:2023avo,EventHorizonTelescope:2019dse}, modelling techniques~\cite{Berti:2005ys,Chapman-Bird:2025xtd}, and recent \ac{GW} measurements~\cite{LIGOScientific:2016aoc,LIGOScientific:2021sio}. 
Katy Clough (Queen Mary University of London) followed with a worked-example approach to numerically solving Einstein's equations, highlighting pitfalls from ignored biases or assumptions~\cite{Baumgarte:2021skc,Aurrekoetxea:2024ypv}. 
Both lectures prompted wide-ranging questions spanning theory and practical methods.  
Jen Gupta (University of Portsmouth) discussed her path into science communication, showing how outreach impacts public perception and how innovative tools like 3D printing can expand accessibility~\cite{Bonne:2018,Middleton:2024ytu}. 
Mikhail Korobko (University of Hamburg) then offered strategies for navigating large conferences, from goal-setting to mental health breaks and effective follow-ups. 
Christopher Berry's (University of Glasgow) session on academic writing outlined techniques for structuring papers, keeping text concise and precise~\cite{Stevance:2021}, and streamlining with \LaTeX{} macros. 
The day closed with Bianca Dittrich (Perimeter Institute and University of Waterloo) on quantum gravity approaches at the Planck scale~\cite{Carlip:2022pyh}, and John Veitch (University of Glasgow) on \ac{GW} data analysis~\cite{Chatziioannou:2024hju,LIGOScientific:2025yae}, demonstrating Bayesian methods, population studies, and open challenges for next-generation detectors.  

Day 2 maintained a similar balance between different topics. 
Antonio Carillo (University College Dublin) reviewed electromagnetic signatures of relativistic systems~\cite{Metzger:2019zeh,Vigliano:2024key,Gezari:2021bmb}, while Andrew Taylor (University of Edinburgh) connected cosmological surveys to gravitational science~\cite{Freedman:2021ahq,Kilo-DegreeSurvey:2023gfr,Chiu:2024ptq,DESI:2024mwx}, highlighting how future observatories could challenge the standard model~\cite{Euclid:2019clj}. 
Daniela Doneva (Universität Tübingen) argued for modified gravity in the strong-field regime~\cite{Berti:2015itd,Freire:2024adf,Yazadjiev:2025ezx}, followed by Elisa Maggio (Max Planck Institute for Gravitational Physics) on observational tests of general relativity~\cite{LIGOScientific:2016lio,LIGOScientific:2021sio} and pitfalls of false deviations from it~\cite{Gupta:2024gun,Kwok:2021zny}. 
Jonathan Gair (Max Planck Institute for Gravitational Physics) concluded with an overview of \ac{GW} experiments across the spectrum~\cite{LIGOScientific:2025hdt,KAGRA:2013rdx,Kalogera:2021bya,Abac:2025saz,LISA:2024hlh,Seto:2001qf,Baker:2019pnp,Luo:2021qji,Luo:2025sos,Verbiest:2016vem}, stressing the opportunities (and challenges) posed by expanding data and new detector regimes.  

\subsection{Career panels}

The panel sessions brought together voices from different career stages. 
More senior panellists shared broad perspectives, while more junior members provided recent, practical insights.  

The Academia panel combined early-career and senior researchers. 
Anna Green (Maastricht University) and Simone Mastrogiovanni (Istituto Nazionale di Fisica Nucleare, Rome) had both recently taken up faculty roles, while Christiana Pantelidou (University College Dublin) and Adam Pound (University of Southampton) contributed experience from research fellowships and long-term posts. 
Discussion ranged from career trajectories and work--life balance to the legal and personal challenges of short-term contracts.  

The Outside Academia panel was shaped by practical constraints: in the absence of travel funding, the panellists were primarily invited from University of Glasgow graduates. 
Attendance required either a supportive employer or taking time off, and in either case, this was easier to justify when the time away from work was minimized, and so was easier for those working near Glasgow. 
Scheduling the panel at the end of the day on Friday minimised the time away from work needed to attend. 
Despite the constraints, the line-up was diverse, spanning law, finance, industry, and media. 
Speakers included Rebecca Douglas (Patent Attorney, Hindles), Peter Wakeford (R\&D Engineer, Optos), Jennifer Toher (Senior Architect, JPMorgan Chase), and Erin Macdonald (science advisor to the \emph{Star Trek} franchise). 
Sebastian Khan (Data Scientist, Starling Bank), a Cardiff University alumnus, spoke on continuing \ac{GW}-related research at their current role.  
Discussion focused on career paths, milestones, and timing of transitions. 
Panellists also addressed how far doctoral research influenced their current work and what additional training proved essential as well as work--life balance with their roles. 
A recurring theme was the value of the skills provided by a PhD, and the value of networking.

\section{Reflections}

As a follow-up to the Workshop, the organizing committee prepared a short survey asking about the general impression of the workshop, the balance between the scientific and soft skills sessions, and asking for extended feedback. 
Out of $23$ responses, the majority ($21$) rated the workshop as 4 and 5 out of 5, as shown in Figure~\ref{fig:responses}. 
In the free-text responses, the recurring theme was the diversity of various topics and especially the transferable-skill sessions:
\begin{figure}
    \centering
    \includegraphics[width=0.9\linewidth]{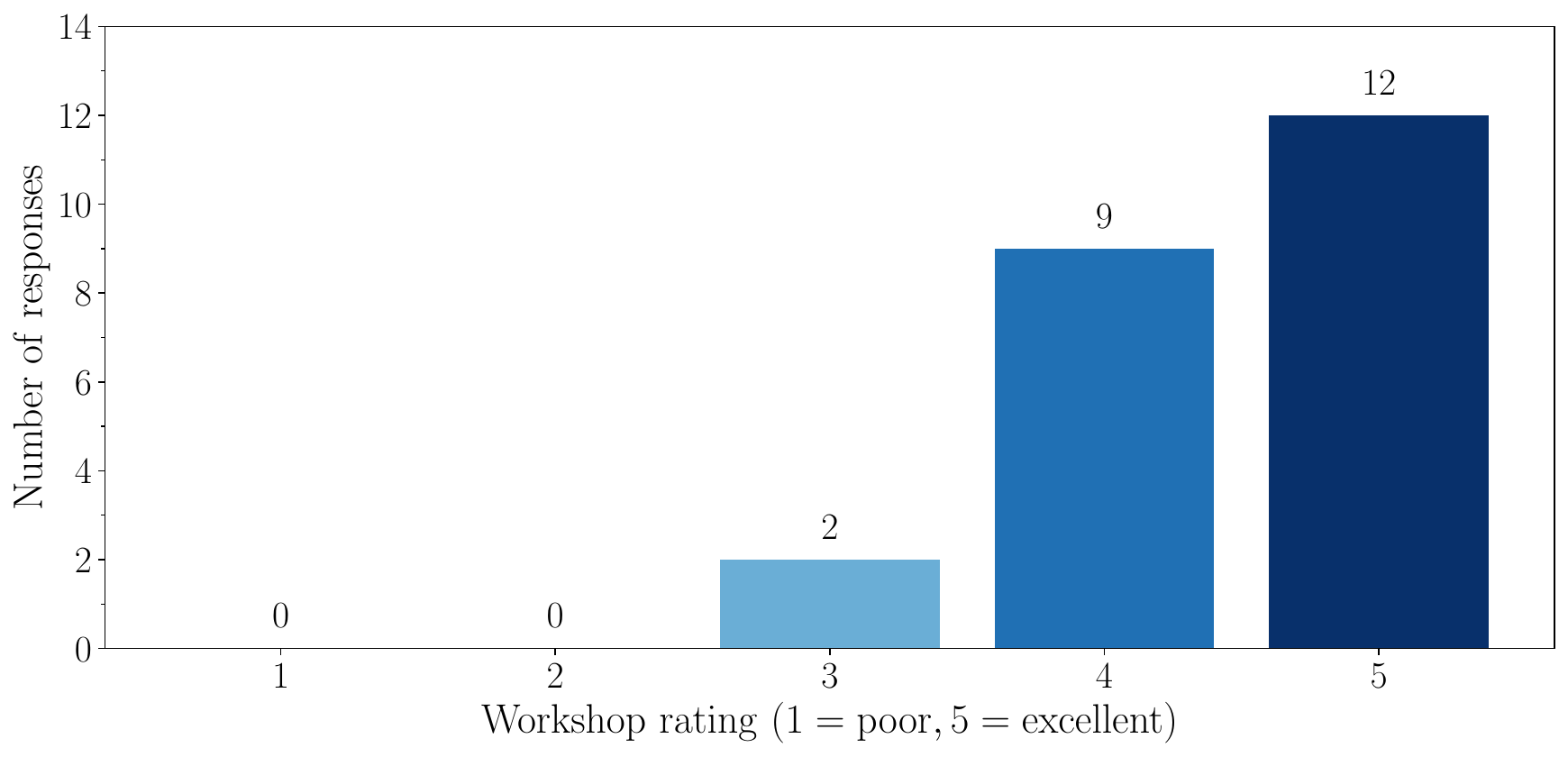}
    \caption{Answers to the survey question ``How would you rate the workshop overall?".
    There were responses from 23 participants out of 141.}
    \label{fig:responses}
\end{figure}
\begin{flushright}
\textit{``Soft skill sessions - a welcome surprise!''}

\textit{``The diversity and informality of sessions was great!''}
\end{flushright}

The participants highlighted the two panel sessions on the career paths in industry and in academia as a chance to expand their knowledge and understanding of possible careers:

\begin{flushright}
\textit{``Beside the science talks that were very useful,\\ 
I really liked hearing about people's history,\\ 
the motivation behind their choices and what they think now about it. \\
This is something that is not common to hear every day!"}

\textit{``I really liked how the workshop gave a clear idea of what it’s like to build a career in this field. \\
It felt like a helpful guide to the future, especially for someone just starting out."}
\end{flushright}

For many, the networking aspect and the chance to meet before the main conference and get to know each other in a more informal atmosphere became the highlight:
\begin{flushright}
\textit{``I also enjoyed the informal networking session at the beginning.\\ 
It made it easier to talk and connect with others during the rest of the days."}

\textit{``Fantastic idea, loved having an intro to topics outside my usual area,\\ 
and the chance to meet others before the large conference started\ldots"}

\textit{``The social events were also great,\\ 
I definitely met more people outside my direct field than I would've at the main conference."}

\textit{``The atmosphere was nice, friendly and relaxed!"}
\end{flushright}

The survey also asked for suggestions for future workshops and a common suggestion was to introduce more hands-on and interactive sessions, both for science and transferable-skills sessions. 
These would also further increase the networking aspect of the workshop. 

Overall, the main theme of the responses was the good balance between the transferable-skill sessions and science introductions, as well as the chance to meet peers.
This matched well with our goals for the Workshop, and highlighted the need for such events within the communities.

\section{Conclusions}

The Early Career Workshop held alongside the \ac{GR}--\ac{Amaldi} meeting in Glasgow provided a rare opportunity for \acp{ECR} to engage with the breadth of gravitational physics and astronomy. 
By combining introductory scientific lectures, transferable-skills sessions, and networking activities, the Workshop created an environment that was both scientifically enriching and socially cohesive. 
The deliberate balance between technical content and professional development ensured participants were better prepared to navigate the main conference and to benefit fully from its diverse programme.  

Feedback from participants confirmed the value of this approach. The scientific lectures offered a structured entry point into areas outside individual specialisations, while the panel sessions gave candid perspectives on career trajectories both within and beyond academia. The networking activities and informal social interactions fostered early connections that carried over into the main meeting, reducing the sense of scale and anonymity that large international conferences can impose.  

The Workshop demonstrated the importance of investing in structured opportunities for \acp{ECR} to build community, develop cross-disciplinary understanding, and acquire skills that extend beyond research. 
In a field that is rapidly expanding and increasingly collaborative, such initiatives strengthen both the individuals involved and the community as a whole. 
Thanks to early engagement with potential sponsors, the Workshop was organised without placing additional financial burden on attendees. 
Future \ac{GR}--\ac{Amaldi} joint meetings should be encouraged to continue to support and expand similar events, ensuring that the next generation of researchers is well prepared to meet the scientific and professional challenges of gravitational science. 

\iftoggle{checklength}{
\section*{Acknowledgments}
The Workshop organising committee graciously thank the sponsors of the meeting: Prof.\ Monica Colpi of the University of Milano-Bicocca; Prof.\ Sheila Rowan of the University of Glasgow; the Royal Astronomical Society; the Institute of Physics Early Career Members, Mathematical \& Theoretical Physics, Astroparticle Physics, and Gravitational Physics Groups; G-Research; Nature Astronomy, and Starling Bank. 
We thank all the speakers for the time and effort they invested in their presentations, and we thank the participants for engaging so wonderfully.
}

\bibliographystyle{iopart-num}
\bibliography{ecr-ref}

\end{document}